\documentclass[
reprint,
floatfix,
aps,
prb,
amsmath,
twocolumn,
superscriptaddress,
amssymb,
tightenlines,
]{revtex4-1}

\pdfoutput=1
\usepackage{amsfonts,amssymb}
\usepackage[subnum]{cases}
\usepackage{mathrsfs}
\usepackage{amsmath}
\usepackage[none]{hyphenat}
\usepackage{hyperref} 
\usepackage{bm} 
\usepackage{natbib}
\usepackage{soul} 
\usepackage{color}
\usepackage{longtable}
\usepackage{ textcomp }
\usepackage{verbatim}
\usepackage{fancyhdr} 
\usepackage{lastpage} 
\usepackage{extramarks} 
\usepackage{graphicx} 
\usepackage{lipsum} 
\usepackage{ amssymb }
\usepackage{slashed}
\usepackage{tikz}
\usepackage{comment}
\usepackage[caption=false,listofformat=parens, subrefformat=parens]{subfig}
\usepackage{ marvosym }
\usepackage{array}

\usepackage{amsthm} 

\newcommand{\abs}[1]{\left| #1 \right|} 
 
\let\baraccent=\= 
\renewcommand{\=}[1]{\stackrel{#1}{=}} 

\theoremstyle{definition}

\theoremstyle{remark}

\newcolumntype{C}[1]{>{\centering\let\newline\\\arraybackslash\hspace{0pt}}m{#1}}

\newcommand{\la}{\left <}
\newcommand{\ra}{\right >}

\newcommand{\nd}{n}

\begin{document}
	
\title{Excess electron screening of remote donors and mobility in modern GaAs/AlGaAs heterostructures}
\author{M. Sammon} 
\email[Corresponding author: ]{sammo017@umn.edu} 
\affiliation{School of Physics and Astronomy, University of Minnesota, Minneapolis, MN 55455, USA}
\author{Tianran Chen}
\affiliation{Department of Physics, West Chester University, West Chester, PA 19383 USA}
\author{B. I. Shklovskii} 
\affiliation{School of Physics and Astronomy, University of Minnesota, Minneapolis, MN 55455, USA}

\received{\today}

\begin{abstract}
	In modern GaAs/Al$_x$Ga$_{1-x}$As heterostructures with record high mobilities, a two-dimensional electron gas (2DEG) in a quantum well is provided by two remote donor $\delta$-layers placed on both sides of the well. Each $\delta$-layer is located within a narrow GaAs layer, flanked by narrow AlAs layers which capture excess electrons from donors but leave each of them localized in a compact dipole atom with a donor. Still excess electrons can hop between host donors to minimize their Coulomb energy. As a result they screen the random potential of donors dramatically. We numerically model the pseudoground state of excess electrons at a fraction $f$ of filled donors and find both the mobility and the quantum mobility limited by scattering on remote donors as universal functions of $f$. We repeat our simulations for devices with additional disorder such as interface roughness of the doping layers, and find the quantum mobility is consistent with measured values. Thus, in order to increase the quantum mobility this additional disorder should be minimized.
\end{abstract}

\maketitle

\maketitle

Modern GaAs/Al$_x$Ga$_{1-x}$As heterostructures with an ultra-high mobility two-dimensional electron gas (2DEG) are the result of spectacular progress in molecular beam epitaxy.\cite{stormer:1979,pfeiffer:1989,umansky:1997,pfeiffer:2003,umansky:2009,UmanskyReview,ManfraReview,reichl:2014,gardner:2016} 
An increase of the electron mobility by nearly 4 orders of magnitude over the last several decades lead to important discoveries, including odd-\cite{tsui:1982b} and even-\cite{willett:1987} denominator fractional quantum Hall effects and stripe and bubble phases.\cite{koulakov:1996,lilly:1999a,du:1999}

\begin{figure}[b]
	\begin{subfloat}
		{
			\includegraphics[width=\linewidth]{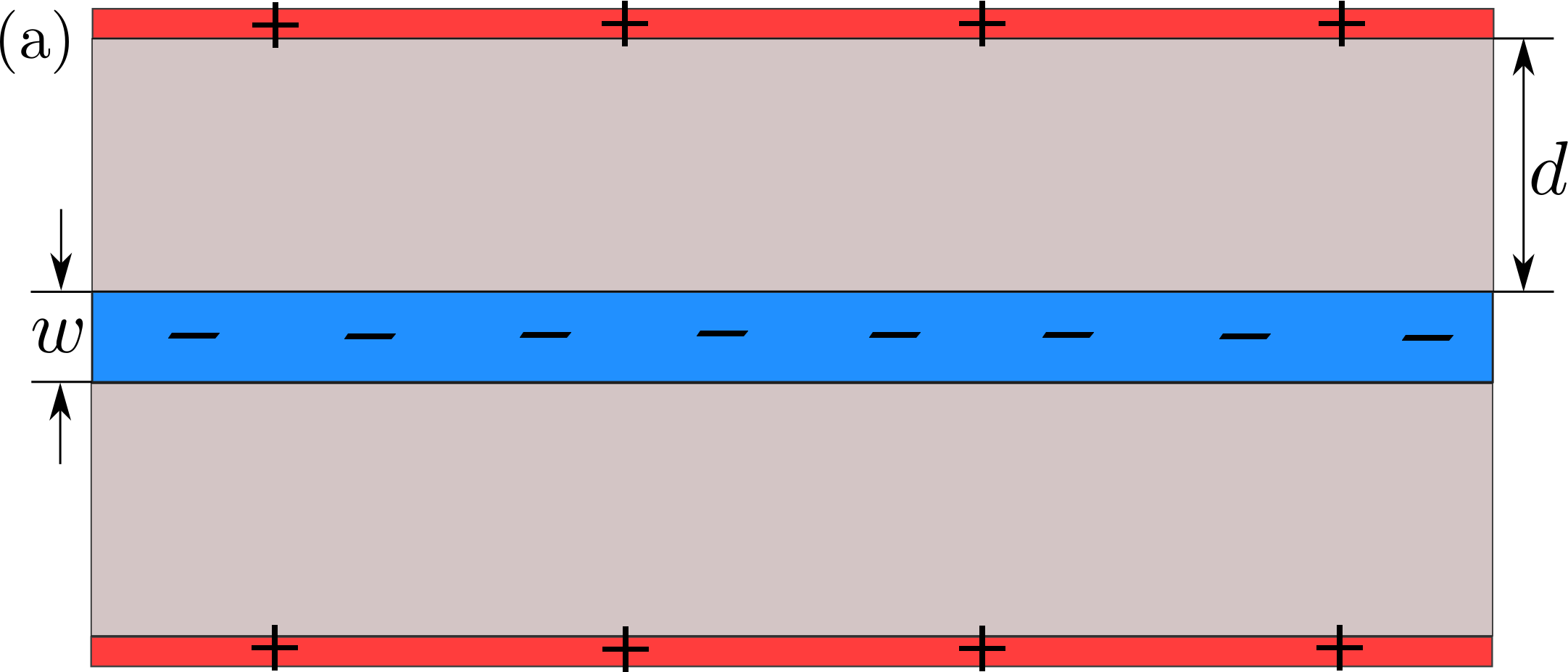}
			\label{fig:device}
		}
	\end{subfloat}
	\begin{subfloat}
		{
			\includegraphics[width=\linewidth]{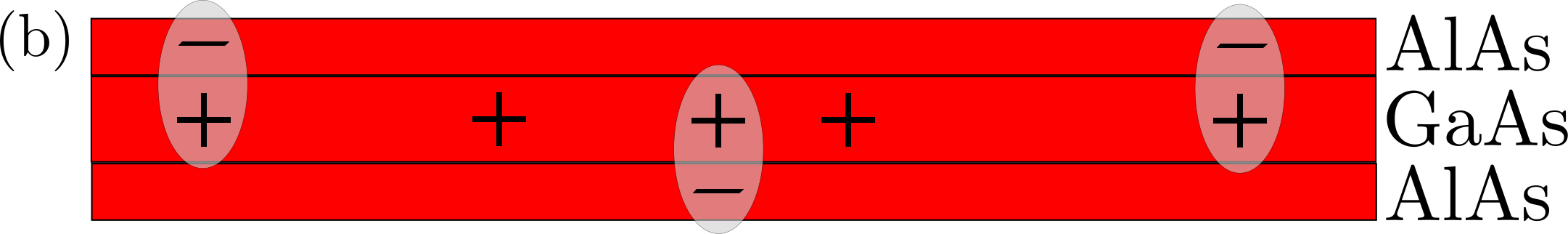}
			\label{fig:dopingwell}
		}
	\end{subfloat}
	\caption{(Color online) (a) A schematic view of a modern GaAs/Al$_x$Ga$_{1-x}$As heterostructure. 
		The 2DEG (shown in blue) resides in a GaAs well of thickness $w$ and is provided by two remote donor layers (shown in red) separated by Al$_x$Ga$_{1-x}$As barriers of thickness $d$ (shown in gray).
		Here, $-$ and $+$ represent negative and positive charges in the 2DEG and the remote donor layers, respectively.
		(b) An enlarged view of a small section of the remote donor layer at a filling fraction $f \simeq 0.6$. 
		Excess electrons ($-$) in AlAs form compact dipoles (ellipses) with the nearest donors ($+$) in GaAs. 
		Empty donors (also shown by $+$) alternate with compact dipoles due to Coulomb repulsion between the excess electrons. 
		Only empty donors are shown in Fig.\,1(a). }
\end{figure}

A typical modern GaAs/Al$_x$Ga$_{1-x}$As heterostructure, schematically shown in Fig.\,\subref{fig:device}, consists of a GaAs quantum well of width $w = 30$ nm. A 2DEG with a concentration $n_e \simeq 3 \times10^{11}$ cm$^{-2}$ is provided to this well by two remote donor layers symmetrically positioned at distances $d\simeq 70-85$ nm from the edge of the well.

These devices have a sophisticated design which substantially reduces electron scattering.\cite{UmanskyReview,ManfraReview} 
As shown in Fig.\,\subref{fig:dopingwell}, each remote donor layer consists of a narrow 3 nm GaAs quantum well, which is doped in the middle by a $\delta$-layer of Si donors with a typical concentration $\nd\sim  10^{12}$ cm$^{-2}$. This layer is surrounded by two AlAs layers of width of 2 nm. 
For these widths of the AlAs and GaAs layers, electrons which are not transferred to the 2DEG (excess electrons) are stored in the AlAs side wells because the relevant effective mass in AlAs is much larger than in GaAs. 
Each excess electron pairs with a donor in a compact dipole atom and is localized, so that its low-temperature parallel-to-2DEG conductance is activated. 
Furthermore, excess electrons hop between donors, minimizing their Coulomb energy; this leads to significant correlations in the positions of charged donors~\cite{Efros1,Dohler,Buks1994,Suski1994,Shikler1997,DasSarma2015} and thus to a dramatic reduction of RD scattering. 
In our recent paper\cite{SammonPhysRevMat} we call this redistribution of electrons excess electron screening (EES).
EES is different from the conventional screening by the 2DEG which exists on top of the EES.

\begin{figure}

			\includegraphics[width=\linewidth]{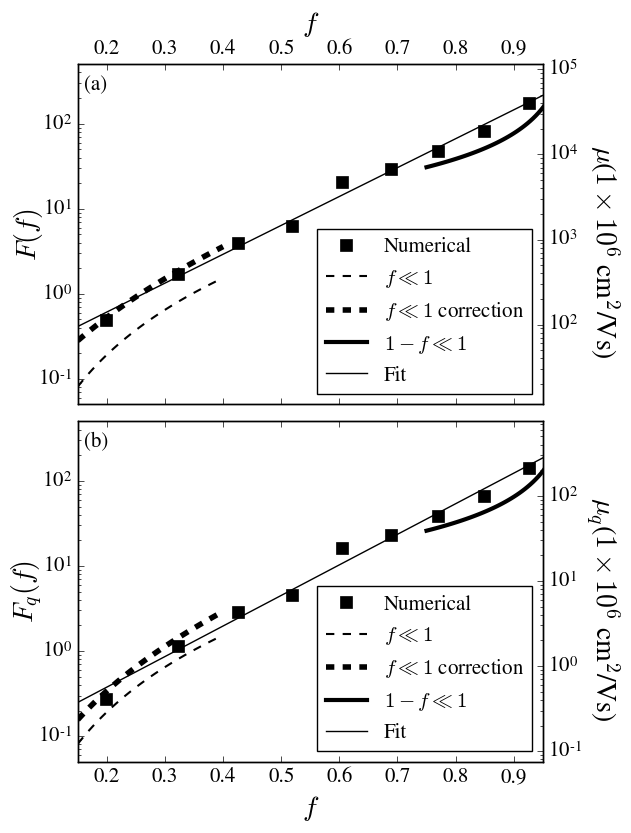}
			\label{fig:2}
	
	\caption{The numerical results (squares) for the dimensionless mobilities $F(f)$ (a) and $F_q(f)$ (b) defined in Eqs.\,(\ref{eq:mobility_final}) and (\ref{eq:quantummobility_final}) plotted on a log-linear scale. Asymptotic estimates\cite{SammonPhysRevMat} Eqs.\,(\ref{eq:F(f)}) and (\ref{eq:F_q(f)}) are shown at $f\ll1$ (thin dashed lines) and at $1-f\ll1$ (solid curves). Improvements to Eqs.\,(\ref{eq:F(f)}) and (\ref{eq:F_q(f)}) at $f\ll1$ for $d_w=9n^{-1/2}$ are shown by the thick dashed lines (see discussion below Eq.\,(\ref{eq:quantummobility})). Best fit Eqs.(\ref{eq:F(f)_fit}) and (\ref{eq:F_q(f)_fit}) are shown by the solid straight lines. Corresponding values of $\mu$ and $\mu_q$ are shown on the right vertical axis for $\mu_0$ and $\mu_{q,0}$ given in Eqs.\,(\ref{eq:mu0}) and (\ref{eq:muq0}) }
	\label{fig:2}

\end{figure}

In Ref.\,[\onlinecite{SammonPhysRevMat}] we presented analytical estimates for the effects of EES on the low temperature mobility $\mu$ and quantum mobility $\mu_q$.\cite{note:0} Here we verify the estimates of Ref.\,[\onlinecite{SammonPhysRevMat}] by numerically modeling EES and calculating both mobilities limited by a single remote donor layer containing donors with concentration $n$ and excess electrons with concentration $fn$. Here $f$ is what we call the donor filling fraction. In the device shown in Fig.\,\subref{fig:device}, neutrality requires that $f=1- n_e/2n$ and $f$ can be varied by changing $n$. In addition, some electrons can be lost to the device surface (not shown) and $f$ can be different for two devices with the same $n$. Thus, for our analysis we treat $f$ as an independent variable. We show below that the mobilities can be written as
\begin{align}
\mu(f) = F(f) \frac{e}{\hbar}k_F^3d_w^5&=F(f)\mu_0,
\label{eq:mobility_final}\\ 
\mu_{q}(f) = F_q(f) \frac{e}{\hbar}k_Fd_w^3&=F_q(f)\mu_{q,0}, \label{eq:quantummobility_final}
\end{align}
where $k_F = (2\pi n_e)^{1/2}$ is the Fermi wavenumber of the 2DEG and $d_w\equiv d+w/2$ is the distance between the midplanes of the quantum well and the remote donor layers. For $n_e=3\times10^{11}$ cm$^{-2}$ and $d_w=90$ nm,  we have
\begin{align}
\label{eq:mu0}
 \mu_0=230\times10^6\, \frac{\mbox{cm}^2}{\mbox{V\,s}}, \\
  \label{eq:muq0}
  \mu_{q,0}=1.5\times10^6\, \frac{\mbox{cm}^2}{\mbox{V\,s}}.
 \end{align}

The dimensionless mobilities $F(f)$ and $F_q(f)$ account for the effects of EES. Their asymptotic expressions at $f\ll1$ and $1-f\ll1$ are\cite{SammonPhysRevMat,note:correction}
\begin{align}
	F(f) & =
	\begin{cases}
		24f^3           &  f\ll1 \\
		7.7(1-f)^{-1} &  1-f\ll1, \\
	\end{cases}\label{eq:F(f)}\\
	F_q(f) & = 
	\begin{cases}
		24f^3 			& f\ll1\\
		6.5(1-f)^{-1} & 1-f\ll1.\label{eq:F_q(f)}\\
	\end{cases}
\end{align}

Eqs.\,(\ref{eq:mobility_final}) and (\ref{eq:quantummobility_final}) are valid only if they predict mobilities larger than the standard values in the presence of $\nd$ donors and no excess electrons ($f=0$),
\cite{price,Gold,dmitriev:2012}
\begin{align}
\mu(0)=\frac{8e}{\pi\hbar}\frac{(k_F d_w)^3}{\nd}\label{eq:mobilityAndo_approx}\,,\\
\mu_q(0)=\frac{2e}{\pi\hbar}\frac{k_Fd_w} {\nd}\label{eq:quantummobilityAndo_approx}\,.
\end{align}
For $n_e=3\times10^{11}$ cm$^{-2}$, $d_w=90$ nm, and $n=10^{12}$ cm$^{-2}$, these mobilities are at least 10 times smaller than the values shown in Fig.\,\ref{fig:2}.

We evaluate $F(f)$ and $F_q(f)$ numerically at all $f$. Our main results are shown by squares in Fig.\,\ref{fig:2}. At $d_w>r_s,k_F^{-1}$, the functions $F(f)$ and $F_q(f)$ should be independent of $d_w$ so that they are universal. Indeed we found that both $F(f)$ and $F_q(f)$ are indistinguishable for $d_w=7$, $9$, and $10$ in units $n^{-1/2}$. For $f \ll 1$ and $1-f \ll 1$ they agree with our Eqs.\,(\ref{eq:F(f)}) and (\ref{eq:F_q(f)}). Best linear fits of the data are given by
\begin{align}
 \log F(f)=3.3f-0.9, \label{eq:F(f)_fit}\\
 \log F_q(f)=3.6f-1.1, \label{eq:F_q(f)_fit}
 \end{align}
  and we see that $F(f)\simeq F_q(f)$ for all $f$. 

We see in Fig.\,\ref{fig:2} that at $f\ll1$, Eq.\,(\ref{eq:F(f)}) is significantly smaller than the numerical results, while Eq.\,(\ref{eq:F_q(f)}) is only slightly smaller. This discrepancy originates from the approximations used in Ref.\,[\onlinecite{SammonPhysRevMat}], where the inverse mobility was calculated to the lowest order in $r_s/d_w$ and made $F(f)$ and $F_q(f)$ universal functions. Restoring the dependence on $r_s/d_w$ significantly improves the agreement at $f\ll1$, as shown by the thick dashed lines in Fig.\,\ref{fig:2}, where $\mu$ and $\mu_q$ were calculated for $d_w=9n^{-1/2}$. For more details see the discussion below Eq.\,(\ref{eq:quantummobility})

Let us now explain how we arrive to these results. First we generate $N=10^4$ randomly positioned donors in a square with side $L$.
Then we find the pseudoground state of the system of $fN$ electrons which occupy $fN$ donors in the presence of a neutralizing uniform background charge with density $-e(1-f)n$, where $n=N/L^2$. All charged donors have oppositely charged point-like images in the 2DEG at the distance $d_w$. We minimize the energy of electrons following the algorithm used in  Refs.~\onlinecite{Bello,Efros1,Dohler} and arrive at the set of charged donor coordinates in a pseudoground state.

The spacial fluctuations of charge is then measured by convolving the charge density of our square with a ``Gaussian envelope". Namely, we calculate the weighted number of charges in our Gaussian envelope centered in the middle of our square at $(0,0)$ as 
\begin{equation}\label{eq:charge_gaussian}
N_R=\sum_i  \exp\left[-\frac{(x_i^2+y_i^2)}{R^2}\right],
\end{equation}
where the sum runs over all charged donors and $R$ is the envelope ``radius". We average $N_R$ and $N_R^2$ over 100 random realizations of our squares for each $f$. Then we find the mean square fluctuation of the number of charged donors in a Gaussian envelope:
\begin{equation}\label{eq:second_moment}
\delta N_R^2=\la N_R^2\ra-\la N_R\ra^2,
\end{equation} 
where $\la...\ra$ denotes averaging over 100 realizations
In the absence of correlations ($f=0$), $\la N_R\ra=\pi R^{2}n$, $\la N_R^2\ra= \pi R^{2}n/2 +(\pi R^{2}n)^2$, and $\delta N_{R}^2=\pi R^{2}n/2$.

 \begin{figure}
	\includegraphics[width=\linewidth]{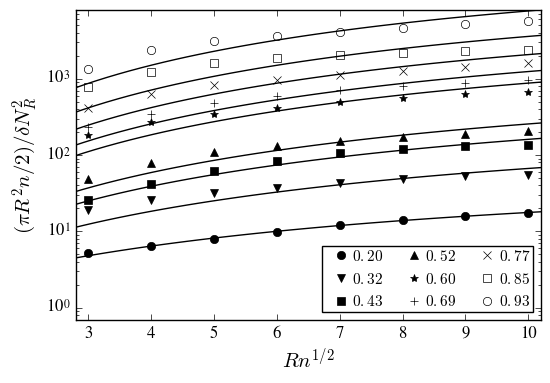}
	\caption{Plots of $(\pi R^{2}n/2)/\delta N_{R}^2$ vs. $Rn^{1/2}$ on a log-linear scale for $0.20\leq f\leq0.93$. Values of $f$ are given in the legend.}
	\label{fig:chfluc}
\end{figure}
The results of our simulation of $\delta N_R^2$ for $0.20\leq f\leq 0.93$ are shown in Fig.\,\ref{fig:chfluc} as the ratio $(\pi R^{2}n/2)/\delta N_{R}^2$ on a logarithmic scale. 
EES reduces $\delta N_{R}^2$ relative to $\pi R^{2}n/2$ dramatically with increasing $f$: $\delta N_R^2\sim 1$ at $f=0.20$ and $\delta N_R^2\sim0.02$ at $f=0.92$. The values of $f$ shown in Fig.\,\ref{fig:chfluc} are measured in the center of the square with the help of the identity $\la N_R\ra=\pi R^{2}n(1-f)$ and are slightly larger than the original $f$ due to the fringe field at the edge of the square. 

 \begin{figure}[t]
 	\includegraphics[width=\linewidth]{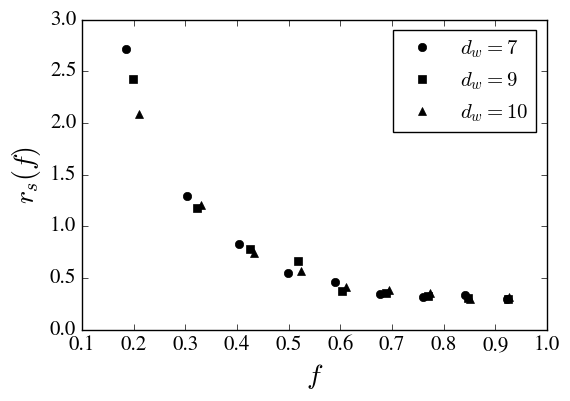}
 	\caption{The effective screening radius $r_s(f)$ in units $n^{-1/2}$ obtained from fits of the numerical simulations for $d_w=7$, $9$, and $10$ in units of $n^{-1/2}$. }
 	\label{fig:rs_vsf}	
 \end{figure}

$\delta N_R^2$ can be related to the correlator of charge density fluctuations $D(\boldsymbol{r},\boldsymbol{r'})=\la n(\boldsymbol{r}) n(\boldsymbol{r'})\ra-\la n(\boldsymbol{r})\ra\la n(\boldsymbol{r'})\ra$ $(\boldsymbol{r}=(x,y)$ is a vector in the $x-y$ plane), where $n(\boldsymbol{r})=\sum_i\delta(\boldsymbol{r_i}-\boldsymbol{r})$. Treating the sum in Eq.\,(\ref{eq:charge_gaussian}) as an integral over $n(\boldsymbol{r})$, Eq.\,(\ref{eq:second_moment}) can be written as
\begin{equation}\label{eq:fluctuations_D(r)}
\delta N_R^2=\int\int D(\boldsymbol{r},\boldsymbol{r'})\exp\left[-\frac{(r^2+r'^2)}{R^2}\right]d^2rd^2r',
\end{equation}
Far from the edges of our square, $D(\boldsymbol{r},\boldsymbol{r'})=D(r-r')$ and we may relate it to its Fourier image $D(q)$ as 
\begin{equation}\label{eq:FT}
 D(r)=\frac{1}{(2\pi)^2}\int D(q)\exp(-i\boldsymbol{q}\cdot\boldsymbol{r})d^2q.
\end{equation}
Combining Eqs.\,(\ref{eq:fluctuations_D(r)}) and (\ref{eq:FT}), we find
\begin{equation}\label{eq:fluctuations_FT}
\delta N_R^2=\frac{R^4}{4}\int D(q)\exp\left[-\frac{(qR)^2}{2}\right]d^2q.
\end{equation}
Below we use,
\begin{equation}\label{eq:D(q)_fit}
D(q)=\frac{(1-f)n(qr_s)^2}{(1+qr_s)^2(1-\exp[-2qd_w])^2},
\end{equation}
to fit Eq.\,(\ref{eq:fluctuations_FT}) and find the screening radius of the excess electrons $r_s(f)$ as a single fitting parameter. Eq.\,(\ref{eq:D(q)_fit}) was used for $f\ll1$ in Ref.\,[\onlinecite{SammonPhysRevMat}] and led to Eqs.\,(\ref{eq:F(f)}) and (\ref{eq:F_q(f)}). Here we have added the additional factor $(1-f)$ because the concentration of charged donors is $(1-f)n$. For $d_w=9n^{-1/2}$ the best fits of our data are shown by the solid lines in Fig.\,\ref{fig:chfluc}. We repeated the simulations for $d_w=7n^{-1/2}$ and $d_w=10n^{-1/2}$ and found the same $r_s(f)$ as shown in Fig.\,\ref{fig:rs_vsf}.

Now the mobilities $\mu$ and $\mu_q$ can be calculated according to 
\begin{gather}\label{eq:mobility}
\mu^{-1}=\frac{ 2\pi\hbar}{e a_B^2}\int\limits_0^{2\pi}\frac{d\theta (1-\cos\theta)e^{-2qd_w}}{(q+q_{TF})^2}D(q)\,,\\ \label{eq:quantummobility}
\mu_{q}^{-1}=\frac{2\pi \hbar}{ea_B^2}\int\limits_0^{2\pi}\frac{d\theta e^{-2qd_w}}{(q+q_{TF})^2}D(q)\,,
\end{gather} 
where $q=2k_F\abs{\sin(\theta/2)}$ is the transferred momentum, $\theta$ is the angle between the initial electron wave vector $\textbf{k}$ and the final wave vector $\textbf{k}+\textbf{q}$, $q_{TF}=2a_B^{-1}$ is the inverse Thomas-Fermi screening radius of the 2DEG, $a_B=\kappa\hbar^2/m^\star e^2 \simeq 10$ nm is the effective Bohr radius in GaAs, and $\kappa$ is the dielectric constant. Using Eqs.\,(\ref{eq:D(q)_fit}), (\ref{eq:mobility}) and (\ref{eq:quantummobility}) with our results for $r_s(f)$ shown in Fig.\,\ref{fig:rs_vsf}, we arrive at $F(f)$ and $F_q(f)$ shown in Fig.\,\ref{fig:2}. 

In Ref.\,[\onlinecite{SammonPhysRevMat}], we used the approximate screening radius $r_s=0.18f^{-3/2}n^{-1/2}$ at $f\ll1$ to calculate $\mu$ and $\mu_q$ using Eqs.\,(\ref{eq:D(q)_fit})-(\ref{eq:quantummobility}). In order to obtain the simple expressions in Eqs.\,(\ref{eq:F(f)}) and (\ref{eq:F_q(f)}), we assumed $r_s\ll d_w$ and set the denominator $(1+qr_s)^{-2}=1$ in Eq.(\ref{eq:D(q)_fit}). In order to improve the agreement with the numerical results in Fig.\,\ref{fig:2}, we have calculated $\mu$ and $\mu_q$ using Eq.\,(\ref{eq:D(q)_fit}) without this approximation for $d_w=9n^{-1/2}$ and the approximate $r_s$ and obtained the thick dashed lines in Fig.\,\ref{fig:2}. For this calculation, we again assumed $k_F^{-1},a_B\ll d_w$, so that the functions $F(f)$ and $F_q(f)$ depend only on $f$ and $nd_w^2$.

 \begin{figure}
 	\includegraphics[width=\linewidth]{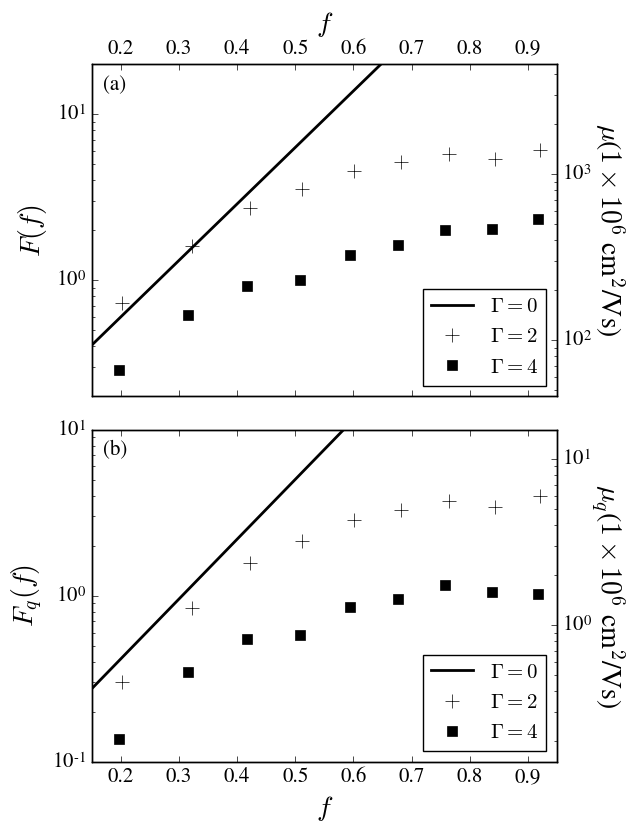}
 	\caption{The universal functions $F(f)$ and $F_q(f)$ obtained from numerical simulations in the presence of additional Gaussian disorder of width $\Gamma$ are shown for $\Gamma=2$ and $\Gamma=4$ in units of $e^2n^{1/2}/\kappa$. The best fit lines for $\Gamma=0$ are given by the solid lines. Corresponding values of $\mu$ and $\mu_q$ are shown on the right vertical axis for $\mu_0$ and $\mu_{q,0}$ given in Eqs.\,(\ref{eq:mu0}) and (\ref{eq:muq0})}
 	\label{fig:disorder}
 \end{figure}

Although our work deals with the same problem as Refs.\,\onlinecite{Efros1} and \onlinecite{Dohler} our results for $\mu$ are different (Refs.\,\onlinecite{Efros1} and \onlinecite{Dohler} did not address $\mu_q$). The difference with Ref.\,\onlinecite{Efros1} is obvious for $ 1-f \ll 1$, where its mobility is much larger than ours. This is because Ref.\,\onlinecite{Efros1} only allowed for large scale fluctuations of donor concentrations, while at $1-f\ll1$ the nearest neighbor disorder which ``melts" the hole Wigner crystal dominates.\cite{SammonPhysRevMat} On the other hand, Ref.\,\onlinecite{Dohler} deals only with a very small spacer $d=10$ nm where EES and 2DEG screening are strongly entangled.

Our results are also different from those of Ref.\,[\onlinecite{DasSarma2015}]. Most of this paper is devoted to Monte-Carlo modeling of correlations of charged donors when electrons must overcome an energy barrier in order to hop to a donor downhill in energy (such as Si donors in AlGaAs forming DX centers). As a result the electron
distribution freezes at some temperature which determines the strength of charged donors correlations.
However, in the modern structures discussed in this paper, electrons see no such barrier for hops between donors that are downhill in energy, and
therefore manage to reach their ground state arrangement on donors which we use to describe correlations.\cite{SammonPhysRevMat}

So far we have dealt only with ideal devices in which the only disorder is the random position of the donors within the $\delta$-layer. In real devices, there are additional types of disorder such as the spreading of the donors throughout the GaAs layer shown in Fig.\,1(b), and roughness of the AlGaAs/AlAs/GaAs interfaces of the remote donor layers.\cite{SammonPhysRevMat} This additional disorder can be quite substantial, for instance the roughness of the AlGaAs/AlAs/GaAs interfaces can shift the quantization energy of the excess electrons by  several $e^2n^{1/2}/\kappa$, where $e^2n^{1/2}/\kappa$ is the scale of the Coulomb interaction. Such large disorder increases $r_s$, weakens EES, and reduces the mobilities. To model this disorder, we added to each donor site a random energy $E$ chosen from a Gaussian distribution $(2\pi)^{-1/2}\Gamma^{-1}\exp[-E^2/(2\Gamma^2)]$. The resulting $F(f)$ and $F_q(f)$ obtained from simulations with $\Gamma=2$ and $\Gamma=4$ in units of $e^2n^{1/2}/\kappa$ are shown in Fig.\,\ref{fig:disorder} along with the best fit results for $\Gamma=0$. Due to increased fluctuations of the results for $\Gamma=2,\,4$, we averaged over 400 realizations of a 100x100 square for both $\Gamma$.  We see that at small $f$ the difference between the mobilities for $\Gamma=2,\,4$ and $\Gamma=0$ is small. However at $f\geq0.4$ the growth of mobilities with increasing $f$ slows and eventually saturates. 
For $\Gamma=4$, and for $n_e=3\times 10^{11}$ cm$^{-2}$ and $d_w=90$ nm, we find that $\mu_q$ saturates at a level comparable to the highest measured values of $1-2\times10^{6}$ cm$^{2}$V$^{-1}$s$^{-1}$,\cite{umansky:2009,Shi(2017)} while $\mu$ is still 10 times larger than the largest experimental values. On the other hand, background impurities may limit $\mu_q$ at the same level.\cite{SammonPhysRevMat} This suggests that the improvement of $\mu_q$ in record samples requires the minimization of this additional disorder together with the reduction of background impurities.

\begin{figure}
 	\includegraphics[width=\linewidth]{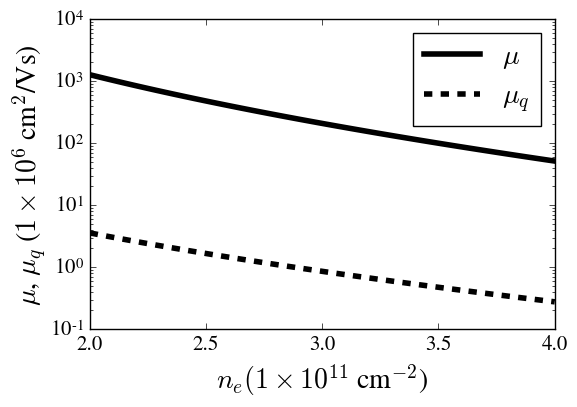}
 	\caption{Mobility $\mu$ and quantum mobility $\mu_q$ as functions of $n_e$ plotted on a log-linear scale. Here we assume the mobilities are limited by a single donor layer with $n=10^{12}$ cm$^{-2}$ donors, where $0.6n$ excess electrons have been lost to the surface.  }
 	\label{fig:mobility_density}
 \end{figure}

Finally, let us mention a possible experiment to verify these results. When the distance $d_w$ between the doping layers and the 2DEG is varied, the 2DEG concentration changes as $n_e\propto 1/d_w$.\cite{ManfraReview} This simultaneously changes the filling fraction in a doping layer according to $f=f_0-n_e/2n$, where $1-f_0$ is the fraction of electrons that the top doping layer has lost to the surface. In Fig.\,\ref{fig:mobility_density} we have plotted $\mu$ and $\mu_q$ using Eqs.\,(\ref{eq:F(f)_fit}) and (\ref{eq:F_q(f)_fit}) as functions of the electron concentration $n_e$ for a fixed donor concentration $n=10^{12}$ cm$^{-2}$ and $f_0=0.4$. Power law fits show that $\mu$ decreases with increasing density as $n_e^{-4.6}$, while $\mu_q$ decreases somewhat slower as $n_e^{-3.7}$. Conversely, in the absence of EES and for $n=n_e$, Eqs.\,(\ref{eq:mobilityAndo_approx}) and (\ref{eq:quantummobilityAndo_approx}) predict much weaker dependencies of $\mu\propto n_e^{-2.5}$ and $\mu_q\propto n_e^{-1.5}$. 
 
In conclusion, we have demonstrated the dramatic effects of EES numerically, and have shown that in an ideal device shown in Fig.\,1 both the mobility $\mu$ and the quantum mobility $\mu_q$ increase by orders of magnitude with the filling fraction $f$ in agreement with Ref.\,[\onlinecite{SammonPhysRevMat}]. In realistic devices, additional disorder in the doping layers may limit $\mu_q$ at values consistent with experimental data. Furthermore, background impurities are known to limit $\mu$ and maybe even $\mu_q$.   This means that while the cleaning of the Ga and Al sources should result in an increase in $\mu$,\cite{reichl:2014,LorenAl} an increase in $\mu_q$ may also require better implementation of the doping layers.

We are grateful to M.A. Zudov, M. J. Manfra, L. N. Pfeiffer, and V. Umansky for helpful discussions. M. Sammon was supported primarily by the NSF through the University of Minnesota MRSEC under Award No. DMR-1420013.


\begin{thebibliography}{32}
\bibitem[{not({\natexlab{b}})}]{note:correction}
	\bibinfo{note}{We use numerical coefficients that are slightly corrected in the
		last version of the ArXiv e-print of Ref.\onlinecite{SammonPhysRevMat}}.
\end{thebibliography}

\end{document}